\documentclass[12pt]{article}

\usepackage{amsmath}
\usepackage{amssymb}
\usepackage{latexsym}
\usepackage{graphics}
\usepackage{psfrag,fancyhdr,epsfig}

\begin{document}

\title{Synergistic Gravity and the Role of Resonances in GRS-Inspired Braneworlds }
\author{C. Bogdanos, A. Dimitriadis and K. Tamvakis\\
Physics Department, University of Ioannina, Ioannina GR451 10,
Greece}
\maketitle
\begin{abstract}
We consider 5D braneworld models of quasi-localized gravity 
in which 4D gravity is reproduced at intermediate scales while the extra dimension
opens up at both the very short and the very long distances, where the geometry is flat. Our main interest is the interplay between the zero mode of these models, whenever a
normalizable zero mode exists, and the effects of zero energy graviton resonant modes coming from the contributions of massive KK modes. We first consider a compactified version
of the GRS model and find that quasi-localized gravity is characterized by a scale for which both the resonance and the zero mode have significant contribution to 4D gravity.
Above this scale, gravity is primarily mediated by the zero mode, while the resonance gives only minor corrections. Next, we consider an asymmetric version of the standard
non-compact GRS model, characterized by different cosmological constants on each $AdS$ side. We show that a resonance is present but the asymmetry, through the form of the
 localizing potential, can weaken it, resulting in a shorter lifetime and, thus, in a shorter distance scale for 4D gravity. As a third model exhibiting quasi-localization, we
 consider a version of the GRS model in which the central positive tension brane has been replaced by a configuration of a scalar field propagating in the bulk.

\end{abstract}

\section{Introduction}

It has been sometime now that theories of extra spatial dimensions, mostly motivated by String/M-Theory, 
have become one of the approaches in attempts to go beyond the
Standard Model physics. In phenomenologically viable models, the extra
 dimensions open up at short distances only, while above a certain length scale physics 
is described by an effective four-dimensional theory. Central to these considerations is the phenomenon 
of localization of gravity \cite{RS1,RS2} arising in the case of the Randall-Sundrum model, a 3-brane embedded in a 5D bulk with negative cosmological constant 
and an infinite fifth dimension. In this setup, four-dimensional gravity is
reproduced at large distances, while the effects of the continuum of massive KK modes is negligible at low energies. Nevertheless, it has been shown by Gregory, Rubakov and Sibiryakov that it is possible to device a setup in which 
extra dimensions open up both at very short and at very long distances\cite{GRS1,GRS2}. In this case, 4D gravity is reproduced only at intermediate scales. At very small scales
there are the standard power-law corrections as in the Randall-Sundrum model, while at very large scales, gravity is again modified and becomes higher-dimensional. 
The root of this phenomenon, named {\textit{``quasi-localization"}}, is the removal of the zero-mode and its replacement by a zero-mass resonance of the continuum spectrum.
The range in which Newtonian gravity is valid depends on the lifetime of this resonance. Extensive analysis of the conditions under which quasi-localization occurs has been performed in \cite{CEH1,CEH2, CEHS} in the case of asymptotic flatness and in black hole backgrounds \cite{Seahra:2005wk,Seahra:2005us,Clarkson:2005mg}. Similar effects also appear in the DGP background \cite{DGP}. The rather counter-intuitive phenomenon of quasi-localization in some specific non-asymptotically flat backgrounds was demonstrated in \cite{STZ}.

In the present article we consider the effects of quasi-localization of gravity in the context of 5D
braneworlds, which exhibit asymptotically flat geometry. We will restrict
ourselves to spaces which look like ordinary Minkowski space
away from the origin of the fifth coordinate. Our main interest is the interplay
between the zero mode of such models \cite{GGS} (whenever this zero mode is normalizable)
and the effects of potential resonances, coming from the contributions of the
massive KK modes of the spectrum. First, we consider a compactified version of the GRS model.
 The masses of the KK spectrum, giving rise to a resonance and effective 4D gravity on the brane, 
 depend on the
size of the compactification radius and exhibit the correct
behavior in the limit of infinite radius, reproducing the GRS effects. Since we assume the extra dimension to be finite, there is always a
normalizable zero mode, contrary to what we encounter in ordinary GRS. We show
that, as the resonance becomes enhanced, the strength of the zero mode on the
brane decreases. The scale of quasi-localized gravity of the GRS model is not
anymore the threshold between 4D and 5D gravity on the brane, but it defines
the distance scale for which both the resonance and the zero mode have
significant contribution to 4D gravity. Above this scale, gravity is mediated
primarily by the zero mode and the resonance gives only higher order corrections.
This situation is similar to the bigravity and multigravity scenarios \cite{KMPRS,KMPR,PAP}, but in
our case, the additional contribution to the gravitational potential doesn't
come from an unusually light KK mode, but is the accumulative product of a host
of these. Next, we consider an asymmetric GRS model, which is non-compact and has
different cosmological constants on each side of a positive tension brane.
Since the model is non-compact and asymptotically flat, there is no zero mode in this setup.
Whatever 4D gravitational effects we get will be only due to the presence of a
resonance. The spectrum of the KK modes in this case is derived and the results
are compared to the ones of the symmetric scenario. We show that a resonance is
indeed present, but the asymmetric localizing potential weakens it, resulting in
a shorter lifetime and thus a shorter distance on the brane, where we get 4D
gravity (see also \cite{Padilla,Charmousis:2007ji}). Finally, as a third example of a model exhibiting quasi-localization, we present a non-compact construction,
 where the central brane is replaced by an appropriate
scalar field defect. 

\section{Compactified GRS}

\subsection{Basic Setup}
We start by considering a metric of the general form
\begin{equation}
ds^2  = e^{ - A\left( y \right)} n_{\mu \nu } dx^\mu  dx^\nu   - dy^2 \,,
\label{ymetric}
\end{equation}
where $\mu,\nu=0,1,2,3$ are 4-D indices and we adopt a $(+,-,-,-,-)$ signature for the metric. The setup is pictorially described in Figure 1. 
We consider the fifth dimension, $y$, to be compactified on a circle of radius radius $R$, such that $-\pi R\le y\le \pi R $ and we 
also assume orbifold symmetry for our system, so we can only deal with the domain $0 \le y \le \pi R$. For negative values of $y$, 
the system is just a mirror reflection. As in the GRS model, we position a brane of positive tension $\sigma$ at the fixed point of the orbifold $y=0$ 
and two branes of negative tension $-\sigma/2$ at some position $y_0$. Notice that the total charge in the orbifold is zero \cite{PAP}. 
We assume that between the positive and negative tension branes there is a negative cosmological constant $\Lambda$, so that the space for $\left| y \right| \le y_0$ is $AdS_5$. For $\left|y \right|>  y_0 $, the central brane is screened by the two negative tension branes and we get Minkowski space. To study this setup, it is convenient to perform a change of variables. Instead of the previous metric, we will use the following ansatz
\begin{figure}[t]
 \centering
     \begin{minipage}[c]{0.8\textwidth}
    \centering \includegraphics[width=\textwidth]{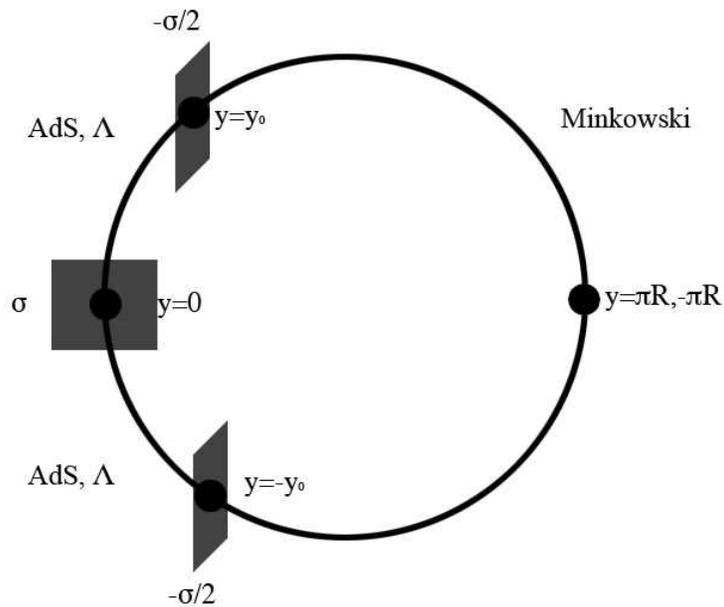}
     \end{minipage}
    \label{figure1}
    \caption{The basic structure of the orbifold, with a positive tension brane located at $y=0$ and two negative tension branes at $y=\pm y_0$.}
 \end{figure}
\begin{equation}
ds^2  = e^{ - A\left( z \right)} \left( {n_{\mu \nu } dx^\mu  dx^\nu   - dz^2 } \right)
\label{zmetric}
\end{equation}
with the same signature. The two coordinate systems are connected by $\frac{{dy}}{{dz}} = e^{ - A/2}$. In order to proceed, we must first re-express the action of the branes in terms of the new metric coordinate $z$, since the definition of the setup was done using the $y$ coordinate. The action of the model with respect to the first metric can be written as
\[
S = S_{gravity}  + S_{branes} 
\]
with the gravitational part being 
\begin{equation}
S_{gravity}  = 2M^3 \int {d^5 x} \sqrt G\, R - \int {d^5 x} \sqrt G \,\Lambda 
\end{equation}
and the action of each brane 
\begin{equation}
S^{(i)} _{brane}  =  - \int {d^4 x}\, \sigma ^{(i)} \,\sqrt { - g\left( y \right)} \, = \, - \int {d^4 x}\int\,dy\,\sigma ^{(i)} \delta \left( {y - y_i } \right)\sqrt { - g\left( {y_i } \right)} 
\label{Sbrane}\,.
\end{equation}
Here $\sigma ^{(i)} $ is the tension of the $i$-th brane located at $y_i$. $G_{MN}$ is the 5D metric and $g_{\mu \nu}$ is the induced 4D metric on each brane. 
We denote the determinant of the induced metric as $g(y)$ to make explicit the dependence on the $y$ coordinate. 
Notice that because of the metric signature we chose, the determinant of the five-dimensional metric $G$ involves a product of four minus signs and hence it is positive. 
However, the induced metric inherits a $(+,-,-,-)$ signature in four dimensions and its determinant is negative. Thus, we write $\sqrt G$ and $\sqrt -g$ respectively.
 We now transform into the $z$ coordinate and get
\[
S^{(i)} _{brane}  =  - \int {d^4 x} \int\,dy\,\sigma ^{(i)} \delta \left( {y - y_i } \right)\sqrt { - g\left( {y_i } \right)}\,  =
 \, - \int {d^4 x}\int\, dz\,\frac{{dy}}{{dz}}\sigma ^{(i)} \delta \left( {y - y_i } \right)\sqrt { - g\left( {y_i } \right)} 
\]
\begin{equation}
 = \, - \int {d^4 x}\int\, dz\,\sigma ^{(i)}\, \delta \left( {z - z_i } \right)\sqrt { - g\left( {y_i } \right)} \,.
\end{equation}
Notice that for the induced metrics we have, $g_{\mu \nu }\left( {y_i } \right)  = e^{ - A\left( {y_i } \right)} n_{\mu \nu }  = 
e^{ - A\left( {z_i } \right)} n_{\mu \nu }  = g_{\mu \nu }\left( z \right) $, so that $\sqrt { - g\left( {y_i } \right)}  = \sqrt { - g\left( {z_i } \right)} $. 
The energy-momentum tensor of each brane $T_{br}^{(i)}$ is derived by varying with respect to the entire 5D metric, 
but we should bear in mind that the action of the brane itself depends only on the induced metric components. Doing so, we obtain
\begin{equation}
T^{(i)} _{br_{{\rm M}{\rm N}} }  =  - \frac{2}{{\sqrt {G\left( z \right)} }}
\,\frac{{\delta S^{(i)} _{brane} }}{{\delta G^{MN}\left( z \right) }} = 
 - \sigma ^{(i)} 
 \delta \left( {z - z_i } \right)
 \frac{{\sqrt { - g\left( {z_i } \right)} }}{{\sqrt {G\left( {z_i } \right)} }}g_{\mu \nu }\left( {z_i } \right) 
 \delta _M^\mu  \delta _N^\nu \,. 
\end{equation}
Since ${\sqrt { - g\left( {z_i } \right)} }/{\sqrt {G\left( {z_i } \right)} }=e^{{A\left( {z_i } \right)}/2} $, the final expression for the brane energy-momentum tensor is
\begin{equation}
T^{(i)} _{br_{{\rm M}{\rm N}} }  =  - \sigma ^{(i)} \delta \left( {z - z_i } \right)e^{\frac{{A\left( {z_i } \right)}}{2}}\, 
g\left( {z_i } \right)_{\mu \nu } \delta _M^\mu  \delta _N^\nu \,. 
\label{Tbrane}
\end{equation}
We see now that the brane tension gets rescaled according to the warp factor of the extra dimension. 
This is the reason which prevented us from stating the initial setup in terms of the $z$ coordinate in the first place. Now that we know the contributions from the branes in the energy-momentum tensor, we can write down Einstein's equations. Our construction is now completely expressed in terms of the $z$ coordinate. We will soon derive the exact relation which maps $y$ points to their $z$ counterparts. The point $y=0$ is in fact mapped to $z=0$, while there is a point $z_0$ corresponding to $y_0$, where the negative tension brane is located. Similarly, we denote by $z_1$ the position which corresponds to $y=\pi R$. Remember that we also have a cosmological constant $\Lambda$ for $\left| z \right| \le z_0$. Einstein's equations are
\begin{equation}
{\cal G}_{MN}  =  - \frac{T_{MN}}{{4M^3 }} \,\,\,,
\end{equation}
where ${\cal G}_{MN}$ is the Einstein tensor. The minus sign comes from our choice of metric signature. With all these taken into consideration, the equations
which describe our setup are
\[
R_{MN} \, - \frac{1}{2}\,R\,G_{MN} \, = \,
\]
\begin{equation}
\frac{1}{{4M^3 }}\left\{\, \Lambda\, \Theta \left( {z_0  - z} \right)\,G_{MN}\,  +
\,\left( \,\sigma \,\delta \left( z \right)\,e^{\frac{{A\left( 0 \right)}}{2}} \, -
\frac{\sigma }{2}\,\delta \left( {z - z_0 } \right)\,e^{\frac{{A\left( {z_0 }
\right)}}{2}} 
\, \right)\,g_{\mu \nu }\, \delta _M^\mu  \delta _N^\nu \,  \right\}\,.
\end{equation}
The first term inside brackets is the cosmological constant, the second comes from the positive tension brane at $z=0$, 
while the third is the contribution of the negative tension brane at $z=z_0$. Remember that we only solve for positive values of $z$. 
Using our ansatz for the metric we arrive at the equations

\begin{equation}
\frac{3}{2}A'\left( z \right)^2\,  = \, - \frac{\Lambda }{{4M^3 }}e^{ - A\left( z \right)} \Theta \left( {z_0  - z} \right)\,,
\end{equation}
\[
\frac{3}{4}\left( {A'\left( z \right)^2  - 2A''\left( z \right)} \right)\, = \, 
\]
\begin{equation}
\,- \frac{1}{{4M^3 }}e^{ - A\left( z \right)}\, \Lambda\, \Theta \left( {z_0  - z} \right)\, - \frac{1}{{4M^3 }}\sigma \,\delta \left( z \right)e^{ - \frac{{A\left( 0 \right)}}{2}} \, +\, \frac{1}{{8M^3 }}\sigma \,\delta \left( {z - z_0 } \right)\,e^{ - \frac{{A\left( {z_0 } \right)}}{2}} 
\,.
\end{equation}
It is easy to see that the solution to these equations for $\left| z \right| \le z_0$ is the Randall-Sundrum metric, $A\left( z \right) = 2\ln \left( {\left|k  \right|z + 1} \right)$, together with the fine-tuning conditions
\[
\Lambda  =  - 24M^3 k^2 ,\,\,\,\sigma  = 24M^3 k
\]
For the rest of the space, the solution is just the Minkowski metric $A(z)=A(z_0)=2\ln \left( {\left|k  \right|z_0 + 1} \right)$, where the value of $A$ is defined 
from the continuity of the metric at $z=z_0$. Having obtained the solution for the warp factor, it is easy to see that for $\left| z \right| \le z_0$ the 
relation connecting the two coordinates $y$ and $z$ is $z = sign\left( y \right)\frac{{e^{k\left| y \right|}  - 1}}{k}$, while for $\left| z \right|>z_0$, 
$z = \left( {k\left| {z_0 } \right| + 1} \right)y$. 
Using these expressions we can map the points of interest
 \[y = 0 \to z = 0,\,\,\,y = y_0  \to z_0  = \frac{{e^{ky_0 }  - 1}}{k},\,\,\,y = \pi R \to z_1  = \left( {k\left| {z_0 } \right| + 1} \right)\pi R\]

\subsection{Zero mode and Kaluza - Klein spectrum}
We are now in position to study the perturbations around the solution we obtained and derive the localizing potential and the spectrum of the fluctuations. We expect to find a zero mode and a tower of Kaluza-Klein states due to the compact nature of the extra dimension. In terms of the $z$ coordinate, we consider perturbations around the 4D Minkowski metric of the form $h_{\mu \nu } \left( {x,z} \right) = e^{3A\left( z \right)/4} \psi \left( z \right)\tilde h_{\mu \nu } \left( x \right)$, with $n^{\alpha \beta } \partial _\alpha  \partial _\beta  \tilde h_{\mu \nu } \left( x \right) =  - m^2 \tilde h_{\mu \nu } \left( x \right)$, $m$ being the Kaluza-Klein mass of the fluctuations. The Schroedinger-like equation for the transverse wavefunction $\psi(z)$ is
\begin{equation}
 - \frac{{d^2 \psi \left( z \right)}}{{dz^2 }} + V\left( z \right)\psi \left( z \right) = m^2 \psi \left( z \right),\,,\label{SCHR}
 \end{equation}
where
\begin{equation}
V\left( z \right) = \frac{9}{{16}}A'\left( z \right)^2  - \frac{3}{4}A''\left( z \right)\,
\end{equation}
is the localizing potential of the setup. Substituting the solution for the warp factor, we have
\begin{equation}
V\left( z \right) = \frac{{15k^2 }}{{4\left( {k\left| z \right| + 1} \right)^2 }} \Theta(\left| z_0-z \right|)
- 3k\delta \left( z \right) + \frac{{3k}}{2}\frac{1}{{k\left| {z_0 } \right| + 1}}\delta \left( {z - z_0 } \right)\,.
\end{equation} 
The solution of (\ref{SCHR}) for $m=0$ (zero mode) is of the form
\begin{equation}
\psi _0 \left( z \right) = Ne^{ - \frac{3}{4}A\left( z \right)} \,=\,N\,\left\{\begin{array}{cc}
\frac{1}{{\left( {k\left| z \right| + 1} \right)^{3/2} }}\,&\,(\left| z \right| \le z_0)\\
\,&\,\\
\frac{1}{{\left( {k\left| {z_0 } \right| + 1} \right)^{3/2} }}\,&\,(\left| z \right| > z_0)
\end{array}\right.
\end{equation}
with $N$ a normalization constant. The normalization condition for $\psi_0 (z)$ 
 \[
\int\limits_{ - z_1 }^{z_1 } {dz\,\psi _0^2 \left( z \right)  = 2\int\limits_0^{z_0 } {dz\,\psi _0^2 \left( z \right)  + 
2\int\limits_{z_0 }^{z_1 } {dz\,\psi _0^2 \left( z \right)  = 1} } }\,
\] 
fixes the constant $N$ to be
 \begin{equation}
  N^2  = 
 k\left[\,1 \,- \frac{1}{{\left( {kz_0  + 1} \right)^2 }}\, +\, 2\frac{{k\left( {z_1  - z_0 } \right)}}{{\left( {kz_0  + 1} \right)^3 }}\,\right]^{-1}\,.
\end{equation}
The above quantity determines the strength with which the zero mode mediates four-dimensional gravity on the brane and will be discussed in more detail later.
 We now turn to the spectrum of massive KK modes. For $m \ne 0$ the equation of motion has the general solution
 \begin{itemize}
\item $\left| z \right| \le z_0$:
\begin{equation}
\psi _m \left( z \right) = A( m )\sqrt {1 + k\left| z \right|}\, J_2 \left( m( k^{-1} + \left| z \right|) \right)\, + \,B( m )\sqrt {1 + k\left| z \right|} \,
Y_2 \left( m( k^{-1} + \left| z \right|) \right)
\label{approx1}
\end{equation}
\item $\left| z \right| > z_0$:
\begin{equation}
\psi _m \left( z \right) = C\left( m \right)\cos m \left| z \right|+ D\left( m \right)\sin m \left| z \right|
\end{equation}
\end{itemize}
This solution is continuous at $z=0$ and $Z_2$-symmetric. We also have to impose continuity at $z=z_0$, as well as discontinuity conditions for the first derivatives at $z=0$ and $z=z_0$. 
Finally, we demand continuity of the wavefunctions (which is automatically satisfied because of $Z_2$-symmetry) at $z=z_1$. Since no brane is present at this position and we only keep symmetric wavefunctions, in order for the solutions to be smooth at $z=z_1$ we have to impose that the derivative vanishes there. We thus have four equations constraining the four constant factors, encoded in the determinant 
\begin{equation}
\left| {\begin{array}{*{20}c}
   {J_1 \left( {\frac{m}{k}} \right)} & {Y_1 \left( {\frac{m}{k}} \right)} & 0 & 0  \\
   0 & 0 & {-\sin mz_1 } & {\cos mz_1 }  \\
   {\sqrt \lambda  J_2 \left( {\frac{m}{k}\lambda } \right)} & {\sqrt \lambda  Y_2 \left( {\frac{m}{k}\lambda } \right)} & { - \cos mz_0 } & { - \sin mz_0 }  \\
   {\sqrt \lambda  J_1 \left( {\frac{m}{k}\lambda } \right)} & {\sqrt \lambda  Y_1 \left( {\frac{m}{k}\lambda } \right)} & {\sin mz_0 } & { - \cos mz_0 }  \\
\end{array}} \right| = 0
\,\,\,\,.
\end{equation}
where we used the abbreviation $\lambda  = 1 + kz_0$. This equation provides us with the spectrum of masses for the KK modes. The quantization condition is apparently non-trivial. In the limit where $z_0 \to z_1$, where the two negative tension branes merge into one with total charge opposite to the one of the positive tension brane, the usual RS-1 setup is reproduced and we get the corresponding KK masses for that model. In all cases, the mass eigenvalues can be found numerically by solving the equivalent equation
\begin{equation}
\tan mz_1  = \frac{{\alpha  + \beta \tan mz_0 }}{{  \beta  - \alpha \tan mz_0 }}
\label{spectrum1}\,,
\end{equation}
where we have defined the coefficients involving Bessel functions
\begin{equation}
\alpha  = J_1 \left( {\frac{m}{k}} \right)Y_1 \left( {\frac{m}{k}\lambda } \right) - J_1 \left( {\frac{m}{k}\lambda } \right)Y_1 \left( {\frac{m}{k}} \right)\,,
\end{equation}
\begin{equation}
\beta  = J_1 \left( {\frac{m}{k}} \right)Y_2 \left( {\frac{m}{k}\lambda } \right) - J_2 \left( {\frac{m}{k}\lambda } \right)Y_1 \left( {\frac{m}{k}} \right).
\end{equation}
In what follows, we will focus on the case where $z_1 \gg z_0$, i.e. the distance between the branes is much smaller than the size of the compactified dimension. Intuitively, since in this case the three branes tend to collapse and vanish because of their opposite charges, we expect the KK modes to be sine and cosine waves, as the space becomes smooth, without any discontinuities and the mass eigenvalues will be given by the simple relation
\begin{equation}
m_n  = n\frac{\pi }{{z_1 }}\,\,\,\,\,\,\,\,\,n=1,2,3...
\label{KKmasses}
\end{equation}
which corresponds to the usual, equidistant spectrum of KK masses for the Laplacian operator. It is easy to show that the right hand side of (\ref{spectrum1}) tends to zero when $z_0 \to 0$, and thus we reproduce the correct relation (\ref{KKmasses}) for small brane separation. In the following discussion, we will work in the regime where $z_1 \gg z_0$, $k z_0 \gg 1$ and consequently the arguments of the Bessel functions will all be small enough to allow asymptotic expansion. Under this assumptions, the spectrum (\ref{KKmasses}) is a good and convenient approximation to the real eigenvalues and we will use this expression from now on. This choice is also justified by numerical results (see Figure 2).

\begin{figure}[t]
 \centering
 \begin{minipage}[c]{\textwidth}
     \begin{minipage}[c]{0.5\textwidth}
     \includegraphics[width=\textwidth]{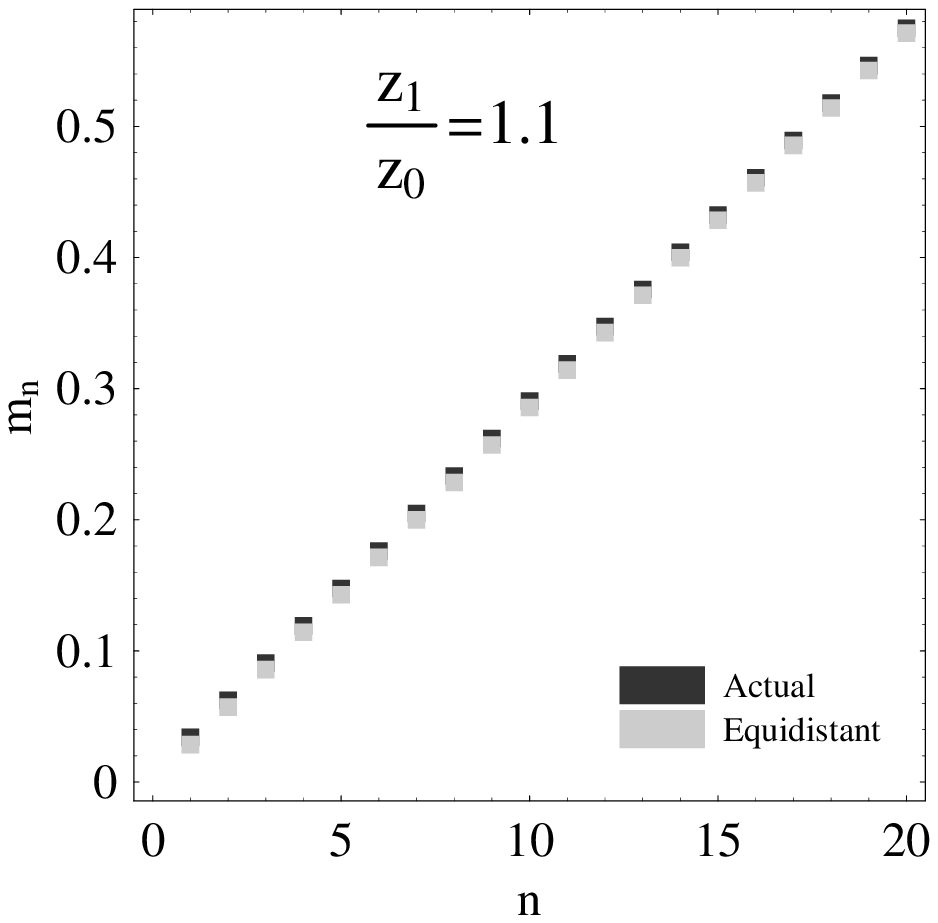}
     \end{minipage}
     \begin{minipage}[c]{0.5\textwidth}
    \includegraphics[width=\textwidth]{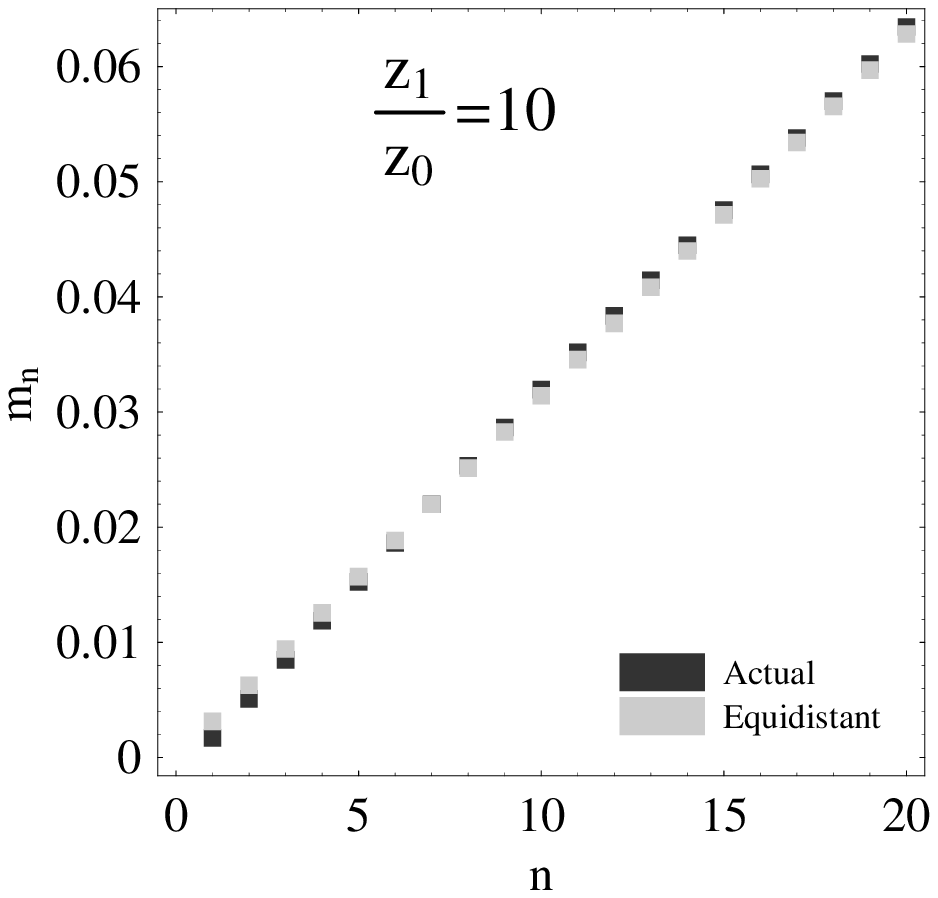}
     \end{minipage}

\end{minipage}
    \label{figure2}
    \caption{Plots of the first 20 Kaluza-Klein eigenvalues against the corresponding equidistant eigenvalues derived from (\ref{KKmasses}), in units of $k$. For large values of the ration $\frac{z_1}{z_0}$, the equidistant spectrum becomes a valid approximation.}
 \end{figure}

Using the junction conditions,
we can now express the three of the four KK coefficients in terms of $A(m)$
\begin{equation}
B\left( n \right) =  - \frac{{J_1 \left( {\frac{{m_n }}{k}} \right)}}{{Y_1 \left( {\frac{{m_n }}{k}} \right)}}A\left( n \right)
\,,\end{equation}
\[
C\left( n \right) = \sqrt \lambda  A\left( n \right)\left\{ {\cos m_n z_0 \left( {J_2 \left( {\frac{{m_n }}{k}\lambda } \right) - \frac{{J_1 \left( {\frac{{m_n }}{k}} \right)}}{{Y_1 \left( {\frac{{m_n }}{k}} \right)}}Y_2 \left( {\frac{{m_n }}{k}\lambda } \right)} \right)} \right.
\]
\begin{equation}
\left. { - \sin m_n z_0 \left( {J_1 \left( {\frac{{m_n }}{k}\lambda } \right) - \frac{{J_1 \left( {\frac{{m_n }}{k}} \right)}}{{Y_1 \left( {\frac{{m_n }}{k}} \right)}}Y_1 \left( {\frac{{m_n }}{k}\lambda } \right)} \right)} \right\}
\,,\end{equation}
\[
D\left( n \right) = \sqrt \lambda  A\left( n \right)\left\{ {\sin m_n z_0 \left( {J_2 \left( {\frac{{m_n }}{k}\lambda } \right) - \frac{{J_1 \left( {\frac{{m_n }}{k}} \right)}}{{Y_1 \left( {\frac{{m_n }}{k}} \right)}}Y_2 \left( {\frac{{m_n }}{k}\lambda } \right)} \right)} \right.
\]
\begin{equation}
\left. { + \cos m_n z_0 \left( {J_1 \left( {\frac{{m_n }}{k}\lambda } \right) - \frac{{J_1 \left( {\frac{{m_n }}{k}} \right)}}{{Y_1 \left( {\frac{{m_n }}{k}} \right)}}Y_1 \left( {\frac{{m_n }}{k}\lambda } \right)} \right)} \right\}
\,.\end{equation}
The remaining overall constant $A(n)$ will be determined by the normalization condition for $\psi_n (z)$
\[
\int\limits_{ - z_1 }^{z_1 } {dz\,\psi _n^2 \left( z \right)  = 2\int\limits_0^{z_0 } {dz\,\psi _n^2 \left( z \right)  + 2\int\limits_{z_0 }^{z_1 } {dz\,\psi _n^2 \left( z \right)  = 1} } } 
\]
 The first of the two integrals involves Bessel functions and cannot be easily performed. 
However, we are interested in the limit where the compactification radius $R$ is very large, so that $z_1 \gg z_0$, that is, 
the distance between the positive and negative tension branes remains finite and much smaller than the extent of the extra dimension. 
In that limit, only the second integral is important and we can extent it to start from zero, since we assume that $z_0$ is small enough. 
We then easily arrive at the normalization condition
\begin{equation}
C\left( n \right)^2  + D\left( n \right)^2  = \frac{1}{{z_1 }}
\end{equation}
which fixes the value of $A(n)$. To derive an expression for $A(n)$, we will make some further simplifying assumptions. 
We will consider that $kz_0 \gg 1$, that is, 
the distance between the branes is much larger than the de Sitter radius. This makes sense, since we want the distance
 scale of 5D short distance gravity ($\frac{1}{k}$) and the quasi-localization scale of effective 4D gravity (which depends on $z_0$) 
 to be well separated. With this assumption, both $m_n z_0$ and $\frac{m_n}{k}$ are much smaller than 1, while $\lambda  = 1 + kz_0  \sim kz_0$. 
 In this case, all the arguments of the Bessel functions will be very small and we can substitute them with their first power series terms.
  Doing so, we arrive at the following expression
\begin{equation}
A\left( m \right)^2  \sim \frac{{\cal A}}{{m^2  + \frac{4}{{z_0 ^2 \left( {kz_0 } \right)^4 }}}}
\,.\end{equation}
where ${\mathcal A} \sim \frac{1}{{z_1 z_0 ^2 }}\left( {\frac{{kz_0 }}{4} + \frac{9}{4}\frac{1}{{\left( {kz_0 } \right)^3 }}} \right)^{ -1}$. Thus the strength of the KK modes at the origin is in this limit
\begin{equation}
\psi _m \left( 0 \right)^2  \sim A\left( m \right)^2 \left( {\frac{1}{{64}}\left( {\frac{m}{k}} \right)^4  - \frac{1}{4}\left( {\frac{m}{k}} \right)^2  + 1} \right) \sim A\left( m \right)^2 
\,.
\end{equation}
as can be deduced from (\ref{approx1}) by imposing our approximations. We see that $A(m)^2$ has the familiar form of a resonance about $m=0$, with a resonance width $ \Delta m \sim 2/z_0 (kz_0)^2 $. 
The characteristic scale at which the resonance decays away from the brane is $r_2  \sim (\Delta m)^{-1} \sim z_0 (kz_0)^2/2$. 
This is exactly what we would expect from the ordinary GRS setup. Thus, the KK spectrum obtained provides us with a resonance in the
 limit of infinite compactification radius, $z_1 \to \infty$. However, an infinite extra dimension is not necessary to get the resonance, 
 which sets in once the KK mode separation becomes small enough, so that the spectrum of masses resembles a continuum. 
 In this case, the minimum of the KK masses approaches zero and we get a contribution in the gravitational potential from the
  lowest lying band of KK modes which reproduces $r^{-1}$ gravity.

\subsection{Static gravitational potential on the brane and synergistic gravity}
We are now in position to calculate the effective 4D potential on the brane due to the spectrum of KK modes and the zero mode.
 As we explained in detail in the previous section, the massive KK spectrum, obeying eq. (\ref{KKmasses}), gives rise to a resonant behaviour once 
 we take the $z_1 \gg z_0$ limit. The existence of the resonance is crucial for the reproduction of 4D gravity from the KK modes on the brane, 
 otherwise their potential will be exponentially suppressed and give a weaker effect, overshadowed by the $r^{-1}$ strength of the zero mode. 
 For the overall static potential between two point particles on the brane we have
\begin{equation}
V\left( r \right) \,=\, V_0 \left( r \right) \,+ \,\sum\limits_m {V_m \left( r \right)\,
=\, \frac{1}{{M^3 }}} \frac{{\psi _0^2 \left( 0 \right) }}{r}\, +
  \,\frac{1}{M^3}\sum\limits_n {\psi _{m_n}^2 \left( 0 \right) \, \frac{{e^{ - m_n r}
 }}{r}}
\,,
\end{equation}
where the first term comes from the zero mode and the second from the sum of KK modes. Using the expression we derived before for the strength of the zero mode 
at the origin, we get for its potential
\begin{equation}
V_0 \left( r \right) \,\sim \,\frac{1}{{M^3 \left[ {1 + 2\frac{{kz_1 }}{{\left( {kz_0 } \right)^3 }}} \right]}}\,\frac{k}{r}\, = 
\,\frac{1}{{M^3 \left( {1 + \frac{{z_1 }}{{r_2 }}} \right)}}\,\frac{k}{r}\,\,\,,
\end{equation}
where we have used the fact that $kz_0\gg1$. We see that once the compactification scale becomes much larger than the characteristic 
distance scale of the resonance, the contribution of the zero mode becomes very small and practically the resonance takes 
over when it comes to effective 4D gravity. In the limit $z_1 \to \infty$ the zero mode becomes non-normalizable and, then, the resonance 
is the sole source of 4D gravity, so we are in the GRS regime.
 The effects of the KK modes on the other hand are represented by the second
term in the potential corresponding essentially to the sum over the Yukawa-like potentials
for all the possible masses of the KK states. Since the separation between modes depends on $\frac{1}{z_1}$, as
$z_1$ gets very large, the spacing between the masses becomes small enough to
turn the sum into an integral over $m$. We should stress that in order to make
this substitution, we should be mindful of the fact that we originally sum over
the KK states, i.e. over the variable $n$, while the integral is over the mass,
the two variables being related by $m_n=n \frac{\pi}{z_1} \Rightarrow 
dm=\frac{\pi}{z_1} dn$. We thus get
\[
V_{KK} \left( r \right) \sim \frac{z_1}{{\pi M^3 }}\int\limits_0^\infty  dm\,e^{ - mr}
\frac{{\psi _m \left( 0 \right)^2 }}{r}
\]
\begin{equation}
 \, =\,\frac{z_1}{{\pi M^3 }}\int\limits_0^\infty  {dm\,\frac{{e^{ - mr} }}{r}}\left(
\frac{{\cal A}}{{m^2  + \Delta m^2 }}\right) \, = \,
 \frac{{{\cal A}z_1 r_2 }}{{\pi M^3 r}}\int\limits_0^\infty  {dx\,\frac{{e^{ - \frac{r}{{r_2 }}x} }}{{x^2  + 1}}} \,.
\end{equation}
Notice that, although the substituted expression for $\psi_m^2 (0)$ is only valid for small masses, the integral is quickly saturated for large $m$ due to the exponential. 
It is thus legitimate to keep the integration limit at infinity. For $r \ll r_2\sim (kz_0)^2 z_0$, i.e. much smaller than the characteristic 
scale at which the resonance decays, we get
\begin{equation}
V_{KK} \left( r \right) \sim \frac{{{\cal A}z_1 r_2 }}{{\pi M^3 r}}\int\limits_0^\infty  \frac{dx}{{x^2  + 1}} \, = \,\frac{{ {\cal A}z_1 r_2 }}{{2M^3 r}}\,.
\end{equation}
We see that the resonance gives a $r^{-1}$ contribution to the static potential. So, for distance scales in the range $\frac{1}{k}\ll r \ll r_2$, 
effective 4D gravity is reproduced by both the zero mode and the resonance coming from the massive KK spectrum. The overall potential is
\begin{equation}
V\left( r \right) \sim \frac{1}{{rM^3 }}\left( \frac{k}{1 + \frac{z_1 }{r_2 }}\, + \,\frac{ {\cal {A}}}{2}z_1 r_2  \right)\,.
\end{equation}
As we infer from this expression, for $z_1\gg r_2$ the resonance dominates. For scales $r\gg r_2$, the only contribution to 4D gravity will be from the zero mode only, 
since the resonance decays and gives 5D corrections of order $r^{-2}$. In the GRS limit the zero mode contribution vanishes and the 4D potential on the brane is 
due to the resonance. Notice that, although the second term in the potential is linear in $z_1$, this is canceled by $\mathcal{A}$, which also has a $z_1^{-1}$ dependence.
 We should also point out that in the GRS limit the massive KK modes should switch to a continuum normalization, not involving $z_1$.
The behavior of this potential can be further understood if we put it into the form
\begin{equation}
V\left( r \right) \sim \frac{(G_{N_0 }  + G_{N_R } )}{r}
\end{equation}
with the definitions for the 4D Newton's constants
\begin{equation}
G_{N_0 }  \sim \frac{k}{M^3\left(1 + \frac{z_1 }{r_2 }\right)},\,\,\,\,\,\,\,\,\, G_{N_R }  \sim \frac{k}{{M^3 }}\,,
\end{equation}
coming from the zero mode and the resonance respectively. 
In deriving $G_{N_R}$ we used that $\mathcal{A} r_2 \sim 2k/z_1$. The ratio of the two gravitational constants in the $k^{-1}\ll r \ll r_2$ regime is
\begin{equation}
\frac{{G_{N_R } }}{{G_{N_0 } }} \sim 1 + \frac{{z_1 }}{{r_2 }}
\,.
\end{equation}
From this relation we infer that the resonance dominates over the zero mode in these distance scales. Only for $r>r_2$, 
where the resonance decays and the relation is no longer valid, does the zero mode take over in mediating 4D gravity on the brane.

\section{An Asymmetric GRS Model}

In this section we shall construct and study an asymmetric extension of the non-compact GRS setup. By asymmetry we mean that the two slices of $AdS_5$ 
around the central, positive tension brane have different cosmological constants, denoted by $\Lambda_+$ and  $\Lambda_-$. The setup resembles Figure 3 and 
the positioning of the branes is done with respect to the $y$-coordinate metric (\ref{ymetric}). The negative tension branes are positioned 
symmetrically around $y=0$, at some distance $y_0$. The action of each brane is given by (\ref{Sbrane}) like before. 
We will switch to $z$-coordinates and the metric (\ref{zmetric}) for convenience in examining the localizing potential. 
The brane energy-momentum tensor takes the form (\ref{Tbrane}). The part of the action due to the cosmological constant is just
 $S_\Lambda   =  - \int {d^5 x} \,\sqrt G\, \Lambda $, where
\begin{equation}
\Lambda\,=\,\left\{\begin{array}{cc}
\Lambda_-\,&\,(-z_0\le z\le0)\\
\,&\,\\
\Lambda_+\,&\,(0<z\le z_0)\\
\,&\,\\
0\,&\,( \left| z \right| >z_0)
\end{array}\right.
\end{equation}
with 
\[
z_0=\frac{1}{k}\left(e^{ky_0 }  - 1\right)\,.
\] 

Einstein's equations for this setup are
\[
R_{MN}  - \frac{1}{2}RG_{MN} \, = \,\frac{\Lambda }{{4M^3 }}\,G_{MN} \,\,+\,
\]
\begin{equation}
 \frac{1}{{4M^3 }}\left( {\sigma _2 \delta \left( {z + z_0 } \right)
 e^{\frac{{A\left( { - z_0 } \right)}}{2}}  + \sigma \delta \left( z \right)e^{\frac{{A\left( 0 \right)}}{2}} 
  + \sigma _1 \delta \left( {z - z_0 } \right)e^{\frac{{A\left( {z_0 } \right)}}{2}} } \right)g_{\mu \nu } \delta _M^\mu  \delta _N^\nu \,, 
\end{equation}
with the cosmological constant $\Lambda$ taking the appropriate value for each interval. Substituting the ansatz for the metric (\ref{zmetric})
 into these equations, we obtain the system of equations
 \begin{equation}
 \frac{3}{2}A'\left( z \right)^2\,  =  \,- \frac{\Lambda }{{4M^3 }}e^{ - A\left( z \right)} \,\,,
 \end{equation}
\[
\frac{3}{4}\left( {A'\left( z \right)^2  - 2A''\left( z \right)} \right) \,=\,  - \frac{\Lambda }{{4M^3 }}e^{ - A\left( z \right)} 
\]
\begin{equation}
 - \frac{1}{{4M^3 }}\left( {\sigma _2 \delta \left( {z + z_0 } \right)e^{-\frac{{A\left( { - z_0 } \right)}}{2}}  + \sigma \delta \left( z \right)e^{-\frac{{A\left( 0 \right)}}{2}}  + \sigma _1 \delta \left( {z - z_0 } \right)e^{-\frac{{A\left( {z_0 } \right)}}{2}} } \right)
\,.\end{equation}

\begin{figure}[t]
 \centering
     \begin{minipage}[c]{0.8\textwidth}
    \centering \includegraphics[width=\textwidth]{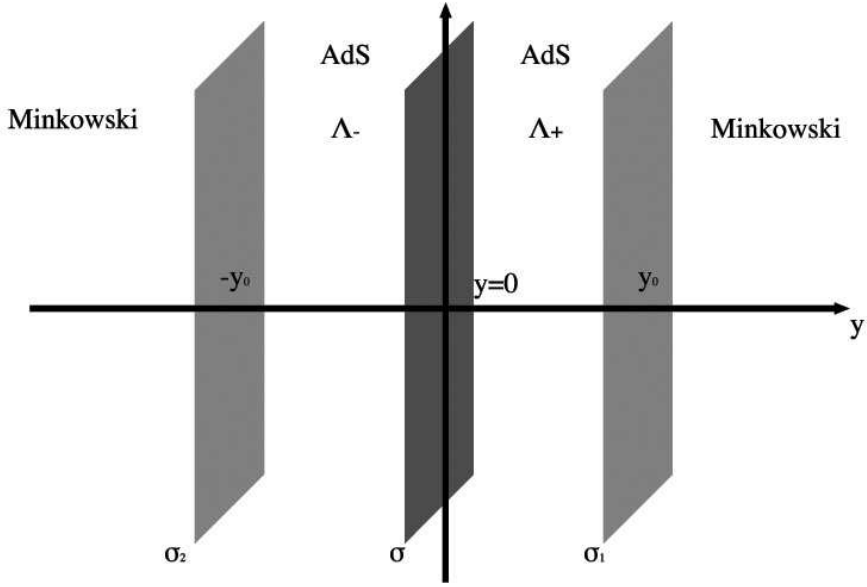}
     \end{minipage}
    \label{figure3}
    \caption{The asymmetric GRS setup}
 \end{figure}
We will assume a solution of the form
\begin{equation}
A\left( z \right) \,= \,\left\{\begin{array}{cc}
2\ln \left( { - k_-  z_0  + 1} \right)\,&\,(z\le-z_0)\\
\,&\,\\
2\ln \left( { - k_-  z + 1} \right)\,&\,(-z_0<z \le 0)\\
\,&\,\\
2\ln \left( { k_+  z + 1} \right)\,&\,(0<z\le z_0)\\
\,&\,\\
2\ln \left( { k_+  z_0 + 1} \right)\,&\,(z> z_0)
\end{array}\right.\,\,.
\end{equation}
Plugging this into the equations of motion we obtain that it is indeed a solution, provided that the following conditions (fine-tunings) hold 
for the brane tensions 
\begin{equation}
\sigma _1  =  - 12M^3 k_ +,\,\,\,\, \sigma _2  =  - 12M^3 k_ -,\,\,\,  \sigma  = 12M^3 \left( {k_ +   + k_ -  }\right)
\end{equation}
and the cosmological constants
\begin{equation}
\Lambda _ +   =  - 24M^3 k_ +  ^2,\,\,\, \Lambda _ -   =  - 24M^3 k_ -  ^2 
\,.\end{equation}
Having obtained the warp factor solution, we proceed to write down the spectrum of excitations for the setup. 
We do this by patching together the original GRS solutions for the KK modes for different k's, so the wavefunctions for the modes will be
\begin{itemize}
\item  $\left| z \right| \le z_0$ 
\begin{equation}
\psi _m \left( z \right) = A_ \pm  \left( m \right)\,\sqrt {1 \pm k_ \pm  z} \,J_2 
\left( m( k_{\pm}^{-1} \pm z ) \right) \,+\, B_ \pm  \left( m \right)\,
\sqrt {1 \pm k_ \pm  z} \,Y_2 \left( m( k_{\pm}^{-1} \pm z) \right)
\end{equation}
\item $\left| z \right| > z_0 $
\begin{equation}
\psi _m \left( z \right) = C_ \pm  \left( m \right)\cos mz \pm D_ \pm  \left( m \right)\sin mz
\end{equation}
\end{itemize}
Notice the $\pm$ sign in the $D$ term, added for later convenience. The setup possesses no normalizable zero mode. 
We will use the junction conditions in order to determine the various constants of the wavefunctions. 
There are three continuity conditions (at $y=0$ and $y=\pm y_0$) and another three equations from the discontinuities 
of the first derivatives at the same points. Thus, in total we have six continuity/discontinuity equations and eight undetermined constants. 
  Imposing the junction conditions we can write all other constants in terms of $A_{\pm}$ 
\[
C_ \pm   = A_ \pm  \sqrt {\lambda _ \pm  } \left[ {\cos mz_0 J_2 \left( {u_ \pm  } \right) - \sin mz_0 J_1 \left( {u_ \pm  } \right)} \right]
 + B_ \pm  \sqrt {\lambda _ \pm  } \left[ {\cos mz_0 Y_2 \left( {u_ \pm  } \right) - \sin mz_0 Y_1 \left( {u_ \pm  } \right)} \right]
\]
\[
D_ \pm   = A_ \pm  \sqrt {\lambda _ \pm  } \left[ {\sin mz_0 J_2 \left( {u_ \pm  } \right) + \cos mz_0 J_1 \left( {u_ \pm  } \right)} \right]
+
B_ \pm  \sqrt {\lambda _ \pm  } \left[ {\sin mz_0 Y_2 \left( {u_ \pm  } \right) + \cos z_0 Y_1 \left( {u_ \pm  } \right)} \right]
\]
with\[
B_ \pm   = \frac{{ \pm Y_1 \left( {\frac{m}{{k_ \mp  }}} \right)\left[ {A_ -  J_2 \left( {\frac{m}{{k_ -  }}} \right) - A_ +  J_2 \left( {\frac{m}{{k_ +  }}}
\right)} \right] - Y_2 \left( {\frac{m}{{k_ \mp  }}} \right)\left[ {A_ -  J_1 \left( {\frac{m}{{k_ -  }}} \right) + A_ +  J_1 \left( {\frac{m}{{k_ +  }}}
\right)} \right]}}{\left[{Y_2 \left( {\frac{m}{{k_ +  }}} \right)Y_1 \left( {\frac{m}{{k_ -  }}} \right) + Y_2 \left( {\frac{m}{{k_ -  }}} \right)Y_1 \left(
{\frac{m}{{k_ +  }}} \right)}\right]}
\]
Note the definitions $u_ \pm   = \frac{m}{{k_ \pm  }}\lambda _ \pm$ and $\lambda _ \pm   = 1 + k_ \pm  z_0$. 

At this point we may adopt a normalization condition that, in the case $\Lambda_+=\Lambda_-$,  would reduce to the standard one. This 
normalization condition reduces to 
\[
\pi\,\left( {C_ +  ^2  + D_ +  ^2 } \right) \,=\,\pi\, \left( {C_ -  ^2  + D_ -  ^2 } \right)\,=\,1\,. 
\]
These two relations suffice to determine all coefficients.

Just as before, we are interested in finding the scale $r_2$ at which the resonance decays, or equivalently, the resonance mass $\Delta m$. To do so, 
we must calculate the strength $\psi _m^2 \left( 0 \right)$ of the KK modes at the origin. 
We are going to make approximations for the arguments of the Bessel functions. Specifically, we assume $k_{\pm}z_0\gg 1$, i.e. 
the distance between the positive and negative tension branes is much greater than the $AdS$ radii on both sides and we also assume
 $mz_0\ll 1  \Rightarrow  \frac{m}{k_{\pm}}\ll 1$. We shall also define the {\textit{asymmetry parameter}} $\eta \,\equiv\, \frac{{k_ +  }}{{k_ -  }}$. 
 Using the above approximations, we end up with the expression for the resonance mass, in the limit of $\eta$ close to unity
\begin{equation}
\Delta m^2  \sim \frac{{8\left( {2 - \eta ^3 \left( {1 - \eta ^3 } \right)} \right)}}{{\left( {\eta  + 1} \right)^2 \left( {k_ +  z_0 } \right)^2 z_0 ^2 }} \sim \frac{{2\left( {2 - \eta ^3 \left( {1 - \eta ^3 } \right)} \right)}}{{\left( {\eta  + 1} \right)^2 }}\Delta m^2 _{GRS} 
\label{asymmetric_m}
\end{equation}
It is evident that the resonance scale depends heavily on the asymmetry parameter $\eta$. As $\eta$ increases, the resonance width becomes bigger and consequently the delocalization distance $r_2$ decreases. Note that the delocalization scale is proportional to
the inverse of the resonance mass. Thus, the resonance becomes weaker as the asymmetry grows, while the greatest resonance scale is achieved in the symmetric setup. We see from (\ref{asymmetric_m}) that the resonance mass of the GRS model is reproduced in the symmetric case, $\eta=1$.

\section{Quasi-localized Gravity on a Thick Brane}

In this section we consider a variant of the ${\cal{Z}}_2$-symmetric GRS model in which the central brane has been replaced by a suitable configuration
of a bulk scalar field \cite{KT}~\cite{KT2}~\cite{Giov}. As in GRS, the $AdS$ space is restricted by two symmetrically placed branes, while the fifth dimension is infinite in extent and
at large distances the space is Minkowski. 
We begin by considering an action of the form 
\begin{equation}
S = \int {d^5 x} \sqrt {  G} \left\{ {2M^3 R + \frac{1}{2}\left( {\nabla \phi } \right)^2  - V\left( \phi  \right)} \right\} - \int {d^4 x} \sqrt { - g} \sigma \left( \phi  \right)
\,,\end{equation}
where we denote the 5D metric by $G_{MN}$ and the 4D metric on the brane by $g_{\mu \nu}$. The first part of the action contains the gravitational and scalar sector, 
while the second is the contribution of a thin brane. The brane tension depends on the scalar field, i.e. there is a scalar potential on the brane.
 This action is written in terms of the metric ({\ref{ymetric}}), which has a $(+,-,-,-,-)$ signature. Let $y_0 \ne 0$ be the location of this thin brane. 
 As it has already been explained, we rewrite the action of the brane and the equations of motion in terms of the metric ({\ref{zmetric}}) of the same signature, which is more
 convenient in the study of the spectrum. Einstein's equations take the form
\[
R_{MN}  - \frac{1}{2}RG_{MN}\,  = \,\frac{1}{{4M^3 }}\left[ {\nabla _M \phi \nabla _N \phi  - G_{MN} \left( {\frac{1}{2}\left( {\nabla \phi } \right)^2 
 - V\left( \phi  \right)} \right)} \right.
\]
\begin{equation}
\left. { \,+ \,G_{MN} \Lambda \,+\, \sigma \delta \left( {z - z_0 } \right)e^{\frac{{A\left( {z_0 } \right)}}{2}} g_{\mu \nu } \delta _M^\mu  \delta _N^\nu  } \right]\,.
\end{equation}
The scalar field equation is
\begin{equation}
\nabla ^2 \phi  + \frac{{dV}}{{d\phi }} = 0\,.
\end{equation}
Notice the sign changes due to the signature used. Substituting the metric ansatz in these equations, we obtain the system
\begin{equation}
A'\left( z \right)^2  - 2A''\left( z \right) =  
- \frac{{e^{ - A\left( z \right)} }}{{3M^3 }}\left( {\Lambda + V} \right) - 
\frac{{e^{ - \frac{{A\left( {z_0 } \right)}}{2}} }}{{3M^3 }}\sigma \delta \left( {z - z_0 } \right) - \frac{1}{{6M^3 }}\phi '^2 
\,,\end{equation}
\begin{equation}
A'\left( z \right)^2  =  - \frac{{e^{ - A\left( z \right)} }}{{6M^3 }}\left( {\Lambda  + V} \right) + \frac{1}{{12M^3 }}\phi '^2 
\,,\end{equation}

\begin{equation}
\frac{1}{2}e^{A\left( z \right)} \left( {3A'\left( z \right)\phi '\left( z \right) -
2\phi
''\left( z \right)} \right) + \frac{{dV}}{{d\phi }} \,=\, -e^{-A(z_0)/2}
\,\sigma'(\phi)\,\delta(z-z_0)\,,
\end{equation}
where $\sigma'(\phi)$ stands for the derivative of the brane-tension with respect to the scalar field at $z=z_0$. 
Only two of these equations are independent. These can be cast in the form
\begin{equation}
\phi '^2  = 3M^3 \left( {A'^2  + 2A''} \right) - \sigma \delta \left( {z - z_0 } \right)e^{ - \frac{A\left( {z_0 } \right)}{2}} 
\end{equation}
and
\begin{equation}
V = \frac{3}{2}M^3 e^A \left( {2A'' - 3A'^2 } \right) - \Lambda - \frac{\sigma }{2}\delta \left( {z - z_0 } \right)e^{\frac{{A\left( {z_0 } \right)}}{2}} 
\,.\end{equation}
 The junction conditions at $z=z_0$ are
\begin{equation}
A'\left( { z_0+ } \right) - A'\left( { z_0- } \right) = \frac{\sigma }{{6M^3 }}e^{ - \frac{{A\left( {z_0 } \right)}}{2}} \,,
\end{equation}
\begin{equation}
\phi '\left( { z_0+ } \right) - \phi '\left( { z_0- } \right) =\sigma'\,e^{ - A\left( {z_0 } \right)/2}
\,.\end{equation}
 Let us now introduce a trial solution for the warp factor function $A(z)$ of the form $A\left(
z \right) =  - \ln \frac{b}{{\left[1 + (kz)^2 \right]}}$. The bulk solution for this setup is
 \begin{equation}
\phi \left( z \right) = \sqrt {12} M^{3/2} \arctan \left( kz \right)
\,,\end{equation}
 \begin{equation}
V = \frac{6k^2 M^3}{b}\left( {\frac{{4\tan ^2 \left( {\frac{\phi }{{\sqrt {12} }M^{3/2}}} \right) - 1}}{{\tan ^2 \left( {\frac{\phi }{{\sqrt {12} }M^{3/2}}} \right) + 1}}} \right)-\Lambda
\,.\end{equation}
This solution is valid for the entire space, even in a smooth scenario without the thin branes present. The junction conditions at $z=z_0$ become
\begin{equation}
\sigma  =  - \frac{{12M^3 k^2 z_0 }}{{\sqrt {b\left( {1 + \left( {kz_0 } \right)^2 } \right)} }}\,,\,\,\,\,\,\,
\sigma ' = -\frac{{\sqrt {12} kM^{3/2} }}{{\sqrt {b\left( {1 + \left( {kz_0 }
\right)^2 } \right)} }}\,.\end{equation}
Notice that for $z_0>0$, the brane tension must be negative. A non-vanishing brane scalar
potential $\sigma'$ is also necessary in order to have a non-trivial solution. We can now
use these junction conditions to patch together the smooth solution with a flat
(Minkowski) space at $z=z_0$. The solution is
\begin{itemize}
\item $z<z_0$:
\begin{equation}
A\left( z \right) =  - \ln \frac{b}{\left[{1 + (kz)^2 }\right]},\,\,\,\,\,\phi \left( z \right) = \sqrt {12} M^{3/2} \arctan \left( kz \right),\,\,\, 
,\end{equation}
\begin{equation}
V = \frac{6k^2 M^3}{b}\left( {\frac{{4\tan ^2 \left( {\frac{\phi }{{\sqrt {12} }M^{3/2}}} \right) - 1}}{{\tan ^2 \left( {\frac{\phi }{{\sqrt {12} }M^{3/2}}} \right) + 1}}} \right)-\Lambda
\,.\end{equation}
\item $z>z_0$:
\begin{equation}
A\left( z \right) =  - \ln \frac{b}{\left[{1 + (kz_0)^2 }\right]},\,\,\, \,\,\phi \left( z \right) = \sqrt {12} M^{3/2} \arctan \left( kz_0 \right),\,\,\, V = -\Lambda
\,.\end{equation}
\end{itemize}
The corresponding junction conditions at $z=-z_0$ are
\begin{equation}
\sigma  =  - \frac{{12M^3 k^2 z_0 }}{{\sqrt {b\left( {1 + \left( {kz_0 } \right)^2 } \right)} }}
\,,\,\,\,\,\,\,
\sigma ' =\,\frac{{\sqrt {12} kM^{3/2} }}{{\sqrt {b\left( {1 + \left( {kz_0 }
\right)^2 } \right)} }}\,.
\end{equation}

Notice that we get the same negative brane tension as before, but $\sigma'$ must now have opposite sign with respect to the brane at $z=z_0$. 
Thus both thin branes must have negative tensions in order to get a solution. This is to be expected.
 The negative tension of these branes is necessary to counteract the positive energy density of the scalar field in the region $-z_0<z<z_0$. 
 The situation is similar to what we encounter in the GRS scenario. The novel property here is that the brane tensions depend on the position of the thin branes.
  This is also not a surprise. The further we position them, the grater the extent of the region where the scalar field is non-trivial will be 
  and the corresponding energy density of the thick brane will grow.

The only apparent problem of the solution is that $V$ may be discontinuous at $z_0$, unless the patch is performed at $z =1/(2k)$. 
However, as we will see, this is before the position where the localizing potential reaches zero for the first time and thus performing 
the patch leaves us with no barrier in the potential. Having a Minkowski space for $z>z_0$ renders the zero mode non-normalizable and, hence,
 gives rise to a proper setup for quasi-localization to set in. 
The localizing potential in the region $z<z_0$ is
\begin{equation}
V_l \left( z \right) = \frac{3k^2}{4} \frac{\left[5(kz)^2  - 2\right]}{{\left[ {1 + (kz)^2 } \right]^2 }}
\,.\end{equation}
The lifetime of the resonance $\Delta m$ is proportional to the transmission factor through the barrier at $m=0$, that is, $\Delta m \sim T(0)$. In the WKB approximation, 
\begin{equation}
\,T\left( m \right) \sim \exp \left( { - 2\int\limits_{z_1 }^{z_0 } {dz\sqrt {V_l \left( z \right) - m^2 } } } \right)\,\,,
\end{equation}
 where $z_1$ is the position at which $V_l=0$, i.e. $z_1  = \pm \sqrt {2/5k}$. The scale of delocalization for gravity, 
 $r_2  \sim (\Delta m)^{-1} \sim T^{ - 1} \left( 0 \right)$, for large values of $kz_0$, becomes
\begin{equation}
r_2  \sim z_0^{\sqrt{15}} 
\end{equation}
For a large enough $z_0$ we can obtain very high values for the scale $r_2$ at which the resonance decays and we get 5D gravity.
 The dependence of the resonance decay scale on the brane distance is similar to the one we obtain from WKB analysis for the GRS model \cite{CEH1}.

We should stress that the results we deduced in this section were based on the validity of the WKB approximation in our setup. 
Another way to demonstrate the presence of the resonance would be to solve exactly for the perturbations of the classical solution we 
obtained above. However, although it is
 straightforward to compute the zero mode wavefunction, 
 solving analytically for the localizing potential is a 
 non-trivial task that deserves further investigation.

\section{Conclusions}

In the present article we considered 5D braneworld models exhibiting the phenomenon of quasi-localized gravity. These models are asymptotically flat, looking like ordinary
Minkowski space at very large distances. We restricted ourselves only in the study of tensor metric perturbations.
We focused on the interplay
between the zero mode of such models (whenever this zero mode is normalizable)
and the effects of potential resonances, coming from the contributions of the
massive KK modes of the spectrum. As a first example, we considered a compactified version of the GRS model.
 Here, as in the case of standard GRS, the massive KK spectrum gives rise to a zero-mass resonance and effective $r^{-1}$ gravitation on the brane, 
 depending on the
size of the compactification radius. The GRS effect is reproduced in the limit of infinite compactification radius. As long as the extra dimension is finite however, 
there is always a normalizable zero mode. We show
that, as the resonance becomes enhanced, the strength of the zero mode on the
brane decreases. This is similar to the results obtained in \cite{GGS} for the asymmetric Randall-Sundrum braneworld. The scale of quasi-localized gravity of the GRS model is not
anymore the threshold between 4D and 5D gravity on the brane, but it defines
the distance scale for which both the resonance and the zero mode have
significant contribution to 4D gravity. Above this scale, gravity is mediated
primarily by the zero mode and the resonance gives only higher order corrections. For distance scales in the range $\frac{1}{k}\ll r \ll r_2\sim (kz_0)^2z_0$,
effective 4D gravity is reproduced by both the zero mode and the resonance coming from the massive KK spectrum. The overall potential is
\[
V\left( r \right) \sim \frac{1}{{rM^3 }}\left( \frac{k}{1 + \frac{z_1 }{r_2 }}\, + \,\frac{ {\cal {A}}}{2}z_1 r_2  \right)\,.
\]
As we infer from this expression, for $z_1\gg r_2$ the resonance dominates. For scales $r\gg r_2$, the only contribution to 4D gravity will be from the zero mode only, 
since the resonance decays and gives 5D corrections of order $r^{-2}$.  The ratio of the two gravitational constants in the $k^{-1}\ll r \ll r_2$ regime
 is 
 \[\frac{{G_{N_R } }}{{G_{N_0 } }} \sim 1 + \frac{{z_1 }}{{r_2 }}\,.\]
 
Next, we considered an asymmetric version of the GRS model in which the two AdS sections have different cosmological constants $\Lambda_+\,\neq\,\Lambda_-$. The model is
non-compact and flat at large distances. Thus, there is no zero mode in this setup.
Whatever 4D gravitational effects we get are only due to the presence of a
resonance. We derived the spectrum of the massive KK modes which, although more complicated than the symmetric case, 
exhibits similar behaviour with a resonance present. We showed that the effects of the resonance, although always present, can be 
weakened by the asymmetry. This is manifest in the localizing potential and can result in 
a shorter lifetime and therefore, a shorter distance quasi-localization distance at which we get 4D
gravity. Finally, as a third example of a model exhibiting quasi-localization, we considered a non-compact construction,
 where the central brane is replaced by an appropriate
scalar field defect. As first observed in \cite{CEH1}, a GRS-like background cannot be reproduced by utilizing only a bulk scalar field. The presence of negative energy density branes is thus imperative and the same pathologies that plague the standard GRS setup \cite{GRS2} remain. Nevertheless, the thick brane model appears viable and is certainly more realistic compared to the simplified thin brane case. As a final note, we should point out that the compactified GRS model may be susceptible to problems involving ghost degrees of freedom, due to the presence of the negative tension branes, as was mentioned in \cite{KMPR,PRZ} (but see also \cite{Neupane:2001st}). A more thorough investigation of scalar perturbations in this context may be necessary to clarify these issues.

{\bf Acknowledgements:} This research was carried out in the framework of the European
Research and Training Network MRTPN-CT-2006 035863-1
(UniverseNet). C. B. acknowledges also an
 {\textit{Onassis Foundation}} fellowship.

\end{document}